\documentclass[12pt, a4paper]{article}

\usepackage[T1]{fontenc}

\usepackage{authblk}
\usepackage{graphicx}
\usepackage{subcaption}
\usepackage[title]{appendix}
\usepackage{amsmath}
\usepackage{lineno}
\usepackage{float}
\usepackage[round]{natbib}
\usepackage[none]{hyphenat}
\usepackage[a4paper, portrait, margin=1in]{geometry}
\usepackage{color}
\usepackage{array}
\usepackage[flushleft]{threeparttable}
\usepackage{setspace}
\usepackage[skip=0.5pt]{caption}
\usepackage[bottom, hang, flushmargin]{footmisc}
\usepackage{pdflscape}

\usepackage[hyphens]{url}

\newcolumntype{L}[1]{>{\raggedright\let\newline\\\arraybackslash\hspace{0pt}}m{#1}}
\newcolumntype{C}[1]{>{\centering\let\newline\\\arraybackslash\hspace{0pt}}m{#1}}
\newcolumntype{R}[1]{>{\raggedleft\let\newline\\\arraybackslash\hspace{0pt}}m{#1}}

\newcommand\blfootnote[1]{%
  \begingroup
  \renewcommand\thefootnote{}\footnote{#1}%
  \addtocounter{footnote}{-1}%
  \endgroup
}

\usepackage{multirow}

\begin{document}
\doublespacing
\onehalfspacing
\title{Levelling Down and the COVID-19 Lockdowns: \\ Uneven Regional Recovery in UK Consumer Spending}
\singlespacing

\author{John Gathergood\footnote{Department of Economics, Nottingham University. john.gathergood@nottingham.ac.uk} \hspace{0.5cm} Fabian Gunzinger\footnote{Warwick Business School, University of Warwick. fabian.gunzinger@warwick.ac.uk} \hspace{0.5cm}  Benedict Guttman-Kenney\footnote{Chicago Booth School of Business, University of Chicago. benedict@chicagobooth.edu} \vspace{0.3cm} \hspace{0.5cm} Edika Quispe-Torreblanca\footnote{Sa\"{i}d Business School, University of Oxford. edika.quispe-torreblanca@sbs.ox.ac.uk} \hspace{0.5cm} Neil Stewart\footnote{Warwick Business School, University of Warwick. neil.stewart@wbs.ac.uk}}

\date{December 21, 2020}

\maketitle

\begin{abstract}

\noindent
We show the recovery in consumer spending in the United Kingdom through the second half of 2020 is unevenly distributed across regions.
We utilise Fable Data: a real-time source of consumption data that is a highly correlated, leading indicator of Bank of England and Office for National Statistics data.
The UK's recovery is heavily weighted towards the ``home counties'' around outer London and the South.
We observe a stark contrast between strong online spending growth while offline spending contracts.
The strongest recovery in spending is seen in online spending in the ``commuter belt'' areas in outer London and the surrounding localities and also in areas of high second home ownership, where working from home (including working from second homes) has significantly displaced the location of spending.
Year-on-year spending growth in November 2020 in localities facing the UK's new tighter ``Tier 3'' restrictions (mostly the midlands and northern areas) was 38.4\% lower compared with areas facing the less restrictive ``Tier 2'' (mostly London and the South).
These patterns had been further exacerbated during November 2020 when a second national lockdown was imposed.
To prevent such COVID-19-driven regional inequalities from becoming persistent we propose governments introduce temporary, regionally-targeted interventions in 2021.
The availability of real-time, regional data enables policymakers to efficiently decide when, where and how to implement such regional interventions and to be able to rapidly evaluate their effectiveness to consider whether to expand, modify or remove them.
\vfill



\end{abstract}

\tiny
\blfootnote{
The views expressed are the authors and do not necessarily reflect the views of Fable Data Limited.
We thank Fable Data Limited for sharing these data for research.
We are grateful to Suraj Gohil, Debbie Mulloy and Fiona Isaac at Fable Data Limited and Lindsey Melynk and Rich Cortez at Chicago Booth for their help facilitating this research. This work is supported by the UK Economic and Social Research Council (ESRC) under grant number ES/V004867/1 `Real-time evaluation of the effects of Covid-19 and policy responses on consumer and small business finances'.}

\normalsize
\doublespacing
\clearpage
\section{Introduction}
Over the last few decades, developed countries have experienced inequality in economic growth at the regional level, with some regions not only experiencing less of the economic boom, but having longer-lasting pain from economic busts.
The UK is one of, if not the most, geographically unequal developed countries.\footnote{\url{https://www.ifs.org.uk/uploads/Green-Budget-2020-Levelling-up-where-and-how.pdf}}
A recent UK government aim is to address such disparities through policies to `level-up' the regions that historically have experienced less benefits arising from globalization and national economic growth.

Yet COVID-19 creates a challenge to such an aim, as both the pandemic and policies undertaken by governments to slow the transmission of the virus can create uneven effects across many dimensions including financial and social capital, age, gender, industry, and locality given differing resilience of individuals and firms to the virus and restrictions.
Understanding these uneven effects is of first order importance for policymakers seeking to address both the short-term economic and social effects of the pandemic and its longer-term implications for exacerbating pre-existing regional inequalities.

In this paper we use granular, real-time data on consumer spending to measure the uneven geography of recession and recovery across the United Kingdom. The granular data is provided by Fable Data, as previously used by \cite{gathergood2020ep} in analysis of the effects of localised lockdowns on local consumer spending.\footnote{More information on Fable Data is available at \url{www.fabledata.com}.} Fable Data record hundreds of millions of transactions on consumer and SME spending across Europe from 2016 onwards and its real-time structure permits research to inform current policymaking. When aggregated, Fable's transaction data provides a highly correlated, leading indicator of official statistics - we find correlation coefficients with Bank of England and Office for National Statistics data of 0.91 and 0.87 respectively but, unlike official statistics, Fable Data is available in real-time: our research access to transactions data is with a one working day lag.

Using these data, we document the uneven geographic impact of COVID-19 on consumer spending in the UK during 2020, shown in total spending and in components of online and offline (i.e., in store) spending. Our headline results is that, while there has been an overall recovery in spending as ``pent-up'' demand has been realised, there is significant geographic variation in the recovery. Aggregate spending recovered from a low of a 29\% year-on-year decline in April 2020 to a 12\% year-on-year growth in October 2020.
Such pent-up demand may be a combination of the lifting of restrictions, consumers becoming more confident to spend given less fear of the virus and improved economic prospects or increased fatigue reducing compliance with restrictions.
We show three key results relating to the geographic variation in recovery.

First, the recovery in consumer spending has been faster in the South and ``home counties'' surrounding London. In contrast, the Midlands, Wales, the North-East and Scotland show the weakest year-on-year growth, or in the latter two cases close to no year-on-year growth at all. Moreover, the faster recovery in the South of England and the Eastern and Western regions is strongly driven by faster growth of online card spending. Notably, within England the fastest year-on-year growth is in the outer-West area of London, the South West and Eastern England -- areas characterised by highly affluent communities and high level of second-home ownership. This suggests that, to a degree, spending growth is strongest in the work from home, or potentially work from second home, areas of the UK.

Second, the speed of recovery has been fastest outside large cities, in commuter-towns and affluent semi-countryside conurbations. We show that the variation in the speed of recovery can also be characterised as differing by types of urban settlements. In particular, the recovery has been fastest and strongest in `business, education and heritage centres' -- such areas are popular domestic tourist destinations and thus this is in line with consumers substituting foreign for domestic holidays.
Recovery was less strong in `countryside living' - predominately rural areas but still noticeably stronger than other, more urban, areas. For more urban areas, London has had a steady recovery whereas `affluent England', `services \& industrial legacy', and `urban settlements' are showing weaker recoveries.

Third, we show that as at the end of November 2020, the point when the second national lockdown was coming to an end to be replaced by the introduction of a revised ``Tier'' system defining levels of restrictions across geographies, the highest Tier areas (known as ``Tier 3'') had experienced a much slower recovery in year-on-year spending, compared with the mid-Tier areas (``Tier 2'').\footnote{There are three Tiers of restrictions applied to geographies in England, Tiers 1-3. We exclude Tier 1 from the analysis as only very few, rural, localities are classed as Tier 1 as of December 2020 (accounting for only 1.3\% of the UK adult population). Scotland is under a different but analogous regime while Wales and Northern Ireland operating under more different approaches.} Tier 2 and Tier 3 areas exhibit similar year-on-year growth rates in card spending in April and July 2020 (the period before this Tier system came into operation). However, by October this pattern diverges, with stronger recovering in overall spending in the Tier 2 areas compared with the Tier 3 areas. This divergence persists through November 2020, with localities facing the UK's new tighter ``Tier 3'' restrictions (mostly the midlands and northern areas) showing 38.4\% lower year-on-year growth in overall spending compared with areas facing the less restrictive ``Tier 2'' (mostly London and the South).
Our results corroborate recent evidence from labour market statistics that the pandemic is levelling down economic activity in the UK, thereby exacerbating regional inequality.

Our study further contributes to a burgeoning literature understanding the economic effects of COVID-19. A succession of studies demonstrate how consumer behaviour has been radically affected by COVID-19 and government policies to mitigate its effects. The first study to do so was \cite{baker2020does} using US fintech data. Following this, Opportunity Insights \citep{chetty2020did,chetty2020real} produced a dashboard using multiple data sources to track regional US consumption behavior alongside other economic indicators.\footnote{\url{https://tracktherecovery.org}}
Beyond the US similar exercises have been carried out to understand household consumption in the early stages of the pandemic -- showing remarkably consistent results \citep{andersen2020consumer,bounie2020consumers,bounie2020consumption,ifs2020,campos2020consumption,carvalho2020tracking,chen2020impact,chronopoulos2020consumer,davenport2020,hodbod20,horvath2020covid,jaravel2020,oconnell2020,surico2020consumption,watanabe2020online}.
Analysis of JP Morgan Chase data \citep{cox2020initial,farrell2020consumption} has described in detail how household balance sheets have changed as a result of the COVID-19 recession and how households have responded to fiscal stimulus.
A variety of studies have examined the effects of the first set of lockdowns on economic behavior and evaluated the degree to which there are trade-offs between policy interventions attempting to contain the virus and economic damage \citep{aum2020covid,beach20201918,barro2020coronavirus,coibion2020cost,correia1918pandemics,cui2020covid,dave2020were,friedson2020did,hacioglu2020distributional, glover2020health,goolsbee2020covid,goolsbee2020fear,guerrieri2020macroeconomic,hall2020trading,lilley2020public,miles2020living,oconnell2020,jones2020optimal,toxvaerd2020equilibrium,wang2020covid}.
A broader  literature has sought to measure regional inequality and understand why it arises and its effects \citep[e.g.][]{milanovic2005half,glaeser2008cities,chetty2018impacts1,chetty2018impacts2,iammarino2019regional,carniero2020long} with recent reports by UK think tanks evaluating the UK government's policy aim to  `level-up' regions.\footnote{
\cite{leunig2008spatial,overman2009case,ludwig2013long,chetty2016effects,geary2016happened,chetty2018opportunity,gal2018reducing,manduca2019contribution,agrawal2020,bhattacharjee2020prospects,carrascal2020uk,davenport2020geography,sensier2020understanding,zymek2020uk}
\url{https://www.ifs.org.uk/uploads/Green-Budget-2020-Levelling-up-where-and-how.pdf} \url{https://www.resolutionfoundation.org/app/uploads/2019/07/Mapping-Gaps.pdf} \\ url{https://www.smf.co.uk/wp-content/uploads/2020/01/Beyond-levelling-up.pdf} \\ \url{https://www.centreforcities.org/wp-content/uploads/2020/02/Why-big-cities-are-crucial-to-levelling-up.pdf} \\ \url{https://www.ippr.org/research/publications/state-of-the-north-2020-21}}
New private sector data sources such as harnessed by \cite{chetty2020did} in response to COVID-19 have been able to reveal in real-time how and why they are developing during this economic and health crisis.

Beyond the topicality and importance of the results, our paper serves to further demonstrate the value of granular, real-time account-level data for economic research. Fable Data contain transaction-by-transaction spending data, updated daily, for a large representative samples of European bank accounts and credit cards, with individual-level and geocode identifiers. As in our earlier paper, we show that these data remain a highly correlated, leading indicator of official statistics -- data which are only available in aggregated form and with many months lag - in contrast to Fable Data which are available in real-time and disaggregated. 
Moreover, these data are applicable to a broad variety of questions in the analysis of individual consumption behavior. They further present a new opportunity for researchers to measure consumption in arguably more reliable ways than using data from financial aggregators (a selected sample of consumers), scanner data (a selected subsample of expenditures) or consumption surveys, which has become less reliable in recent decades and has prompted a variety of initiatives aimed at improving the measurement of consumption \citep[see][]{browning2014measurement, landais2020introduction}.
In current research we are further exploring the potential of these data to analyze retail sectoral impacts.

The ability to measure regional, economic data in real-time using datasets such as Fable Data offers exciting potential to inform when, where and how to target regional policy interventions for evidence-based policymaking.\footnote{To this end the authors have access to a variety of real-time, high-quality private-sector datasets for research to inform policymaking. If you are a data provider interested in joining this collaboration please contact the authors for further details on how to potentially partner in this initiative.}
In particular, the ability to evaluate the causal effects of interventions across a breadth of economic outcomes in real-time provides a cost-effective way for governments to trial interventions in selected regions and quickly consider whether to expand, modify or remove such policies.
This is a more nimble strategy than traditional government approaches of typically applying policies at a national level with limited abilities to assess their impacts.

In the context of this paper, the regional inequalities shown indicate that, in order to `level-up' the historically less productive UK regions longer-term, there is a rationale for trialling short-term interventions to address the 'levelling-down' that has occurred in 2020 as a result of COVID-19.
These could occur in 2021 once the virus outbreaks are under control and vaccinations have been more broadly rolled out.
What could such measures look like? 
Encouraging spending in businesses located in harder hit areas could occur through business rate relief or VAT cuts.
Other measures (e.g. travel vouchers) could try to encourage consumers to visit harder-hit parts of the UK.
Or a less centralised approach would be for national governments to make temporary funding available to local governments in proportion to how adversely they have been impacted by the crisis for those local authorities to spend as they see fit (e.g. council tax reductions or rebates, funding local events or services).
We hope our paper will prompt public discussion on other types of measures that could be feasibly implemented and evaluating their merits.

\section{Data}

\subsection{Consumption Data}

We use consumption data provided by Fable Data Limited as previously used in \cite{gathergood2020ep}, and summarize again here the key features of the data.\footnote{More information on Fable Data is available at \url{www.fabledata.com}.} Fable data record hundreds of millions of transactions on consumer and SME spending across European countries from 2016 onwards.\footnote{Commercial sensitivities mean we do not disclose the exact number of accounts and transactions available in the data.}
Fable's transaction data are anonymized and available in real-time: our research access is with a one working day lag. Fable sources data from a variety of banks and credit card companies: accounts cover both spending on credit cards and inflows and outflows on current (checking) accounts. Data is at the account-level and hence we can follow spending behavior on an individual account over time.\footnote{In cases where one individual has multiple accounts, we cannot link multiple accounts in the data to the individual but can aggregate to a geographic region.} Fable data is similar to recently-available data sets from financial aggregators and service providers, but does not have some of the limitations of other datasets as Fable Data works solely with anonymised datasets, and has sourced data directly from banks and credit card issuers, rather than individual subscribers.\footnote{\cite{baker2018debt} provides validation and application of US financial aggregator data. Financial aggregator data for the UK is widely shared for research purposes by Money Dashboard, a UK-based fintech \citep{chronopoulos2020consumer,davenport2020,ifs2020,surico2020consumption}. \cite{ifs2020} analyse the characteristics of Money Dashboard users.}

For each spending transaction we observe a standard classification merchant category code for the spending type. Fable also produces its own categorizations of spending, utilizing the more granular information it has available from transaction strings. These data also differentiate between online and store-based transactions.

The data has an added feature of containing geo-tags for both the card holder's postcode sector and, where applicable, the address of the store or outlet in which a transaction is made. For each UK account we observe the postcode sector of the cardholder's address. In the UK, postcode sectors are very granular geographies: There are over 11,000 postcode sectors in the UK with each sector containing approximately 3,000 addresses. Where a transaction can be linked to a particular store, the full address of that store is available. Also, where a transaction is of a listed firm, Fable tags merchants to their parent groups and stock market tickers. For this study we focus on transactions denominated in British pounds sterling on UK-based credit card accounts held by consumers.

A feature of transaction-level spending data is that, even in data sets containing large volumes of transactions, the value and count of transactions tend to be highly volatile. Volatility arises across days of the week, weeks of the month, public holidays, and to some extent due to variations in the weather. Hence, to construct daily series for comparison with official statistics, we follow an approach to smooth the transaction volumes over time as used by Opportunity Insights on similar US data \citep{chetty2020real,chetty2020did}: aggregating spending by day at the level of geography of interest, taking a seven day moving average and dividing by the previous year's value.\footnote{For 29 February 2020 we divide by an average of 28 February and 1 March 2019.} We normalize these series setting an index to 1 using the mean value 8 - 28 January 2020. We also construct daily series using a 14 and 28 day moving average in an analogous fashion. Finally, for our geographic spending comparison, we calculate year-on-year monthly growth. We use both year-on-year and post-January 2020 calculations in our regional analysis.

\subsection{Comparison with Official Statistics}

One key advantage of Fable Data is the speed with which it can be made available to researchers and policymakers. This is among many features of the data which make it attractive for research -- the timeliness (it is available the next working day, whereas official statistics are typically available only with a lag of several months), geographic granularity (being available at a lower level than official statistics) and, transaction-level (enabling a more flexible analysis than aggregated official statistics). These data can therefore potentially be used to construct leading indicators for policymakers and enable researchers to answer a broader set of research questions than was previously possible using more traditional data sources.

\begin{figure}[H]
	\centering
	\caption{\textbf{UK Credit Card Spending 2018 - 2020 }}
	\vspace{1cm}
	\begin{tabular}{c}
		\textbf{A. Fable Monthly Data Compared to Bank of England \&} \\
		\textbf{\& Office for National Statistics Data, 2018 - 2020} \\
		{\includegraphics[height=3in]{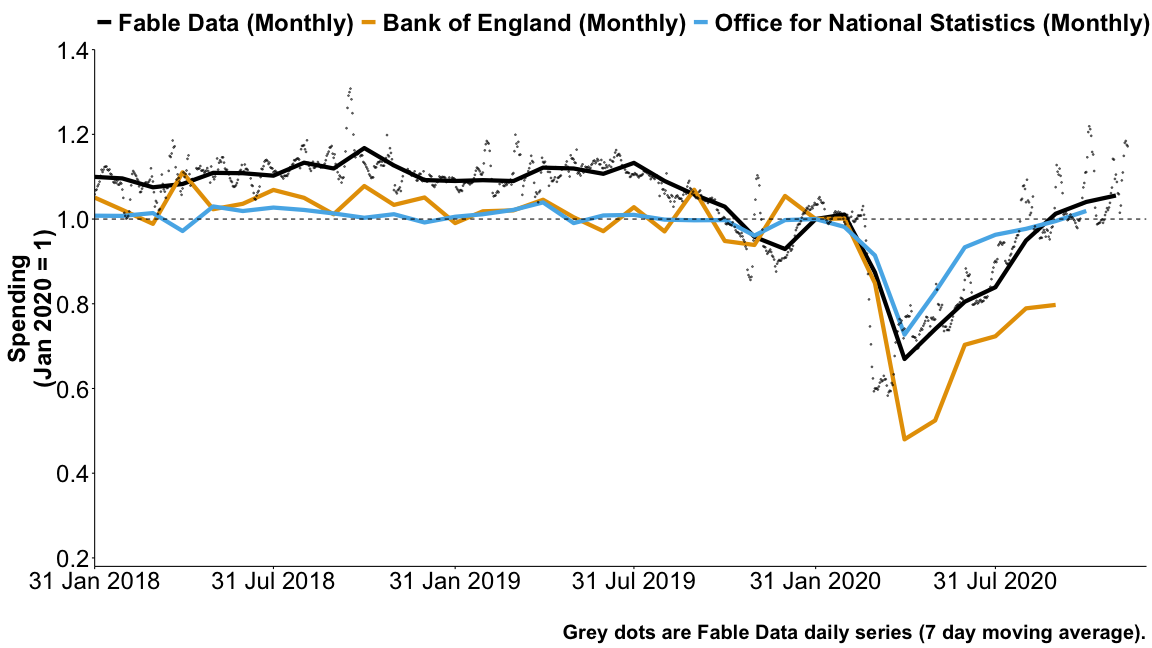}} \\ \\
		\textbf{B. Fable Daily Data, 2020} \\ \textbf{(7, 14, 28 day moving averages and monthly)} \\
		{\includegraphics[height=3in]{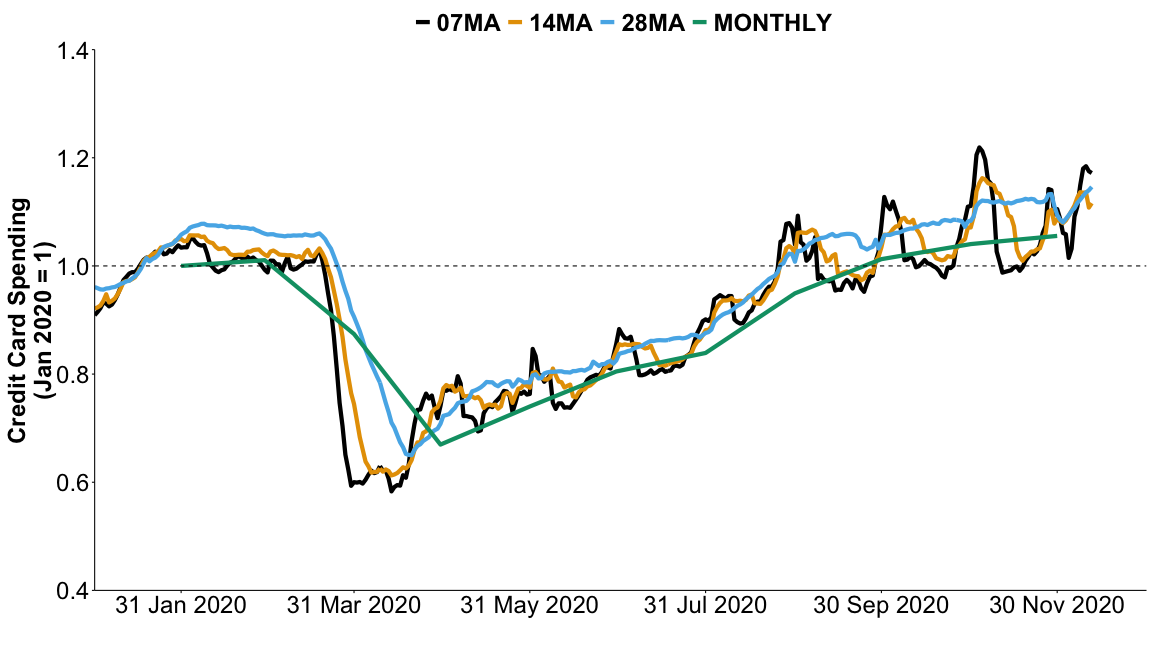}}
	\end{tabular}
	\begin{tablenotes}
		\small
		\item \textit{Notes: Bank of England monthly data is derived from LPMVZQH (monthly gross credit card lending to individuals). Office for National Statistics monthly data is derived from value of all retail sales (average weekly sales for all retailing including automative fuel).
			Fable Data monthly series is indexed to January 2020.
			Fable Data 7,14,28 day moving averages are the daily moving average de-seasoned by taking ratio of the moving average a year prior. Each daily series is then indexed to its moving average 8 - 28 January 2020. Fable Data to 12 December 2020.
		}
	\end{tablenotes}
	\label{fig:boe}
\end{figure}

However, while these features are potentially valuable, their usefulness depends in part on how this data series relates to comprehensive, official data. To explore this, Figure \ref{fig:boe} Panel A (which updates an earlier version of this figure as shown in \cite{gathergood2020ep}), compares the time series of Fable Data UK annual changes in monthly credit card spending to the Bank of England series and shows they are highly correlated: correlations 0.90 (January 2018 to September 2020), 0.87 (January 2019 to September 2020) and 0.90 (January 2020 to September 2020).
Bank of England data is only published in aggregated form monthly and with a lag.
This figure also compares to Office for National Statistics data on the value of retail sales (which is available at a slightly shorter time lag) and similarly shows Fable as a leading indicator of this with the two series to be highly correlated: correlations 0.87 (January 2018 to October 2020), 0.88 (January 2019 to October 2020) and 0.91 (January 2020 to October 2020).

Figure \ref{fig:boe}, Panel B shows Fable data measures for 7, 14, 28 day moving averages -- which can be calculated daily in real-time -- compared to the monthly series (which requires waiting until the month end).
These daily moving averages show the sharp drop in consumption in March 2020 far earlier than the monthly series.
We thus conclude that we can use these data as a reliable real-time predictor of official data and as a reasonable proxy for measuring consumer spending.

On aggregate, we observe a sharp fall in UK credit card spending near the time of the spike in Covid-19 cases and the national lockdown announcement on 23 March 2020 and then a fairly steady recovery from May to August. Through to the first few weeks of December 2020, we observe that the recovery in credit card spending has continued, although not in a sharp ``V-shape'' but instead in a shape resembling a ``tick-shape''. This indicates that, in aggregate, there has been no evident bounce-back to account for lost spending through the second and third quarters of 2020.
We also note that these patterns of spending are for UK residents who, without international travel, are spending more time (and therefore money) domestically but for considering the broader economy and particular sectors within it such growth in domestic spending is unlikely to be sufficient to compensate for the lack of spending by tourists to the UK.

\section{Results}

\subsection{Aggregate Card Spending}

Table \ref{tab:aggregate} summarises year-on-year changes in overall card spending and then disaggregate this into offline and online spending for April, July, October, and November 2020. In April the UK faced the first -- and tightest -- national lockdown including closure of all non-essential shops, a requirement or workers to work from home wherever possible, and limitations on exercise and leisure outside the home. Overall, card spending fell by more than one-quarter year-on-year, with a drop of over 40\% in offline spending. While online spending increased, the increase year-on-year is a modest 2.4\%, partly due to stock-piling prior to the onset of the first national lockdown and partly due to the limited capacity of retailers to increase the distribution of goods and services through online channels.

Through the two subsequent quarters of 2020, we see a recovery in card spending, notably dominated by large year-on-year increases in online spending. Through July, spending recovered to approximately 10\% down year-on-year, but rebounded to approximately 10\% up year-on-year as restrictions eased and some ``pent-up'' demand was realised. The modest recovery in offline spend, which was still 4\% down in October, was greatly outstripped by a surge in online spend, which was 40\% up in October, year-on-year. A series of local restrictions in the late summer and early autumn did not appear to result in large declines in spending \citep{gathergood2020ep}. The second national lockdown in November (which saw some restrictions on non-essential shops, but not to the same extent as in April and May 2020) saw a further dip in offline spending, causing a fall of approximately 12\% year-on-year, while online spending continued to grow, at at year-on-year rate of more than 50\%. This contributed to a steady year-on-year increase in overall spending in November.

\subsection{Regional Variation}

Our focus in this paper is on the regional variation underlying the national trend. 
Table \ref{tab:aggregate_regions} repeats Table \ref{tab:aggregate} showing year-on-year changes separately for Northern Ireland, Scotland, and Wales.
This reveals how total and offline spending in Wales was noticeably lower in November 2020 than Scotland but both regions had similar online spending growth. 

The flexibility of Fable's transaction-level data enables us to decide the most suitable level of geography to examine and measures to construct.
To illustrate this, we calculate our two measures of consumer spending (year-on-year changes, and normalised to Jan 2020) at the mid-level geography of the UK, known as the Nomenclature of Territorial Units for Statistics Second Tier, or NUTS 2.

Figure \ref{fig:map_april} illustrates the year-on-year change in overall card spending (Panel A), offline spending (Panel B), and Online spending (Panel C). The colour shades have different scales on each panel but red indicates a decline in year-on-year spending, while the colour shades in blue indicate a growth in year-on-year spending and whiter shades indicate being closer to no year-on-year growth. The figures illustrate the aggregate trend in card spending, with a national decline in overall spending across the United Kingdom, driven by a particularly strong decline in offline spending and moderate growth in online spending.

Figure \ref{fig:map_october} reproduces the same illustrations for October 2020, a point at which the UK had emerged from the end of first national lockdown and instead briefly instigated three tiers of restrictions, applied at the local level. The figures show regional heterogeneity in the speed of recovery in spending. Overall, card spending shows the strongest year-on-year growth in the South of England and the Eastern and Western regions, while the Midlands, Wales, the North-East, and Scotland show weak year-on-year growth, or in the latter two cases close to no year-on-year growth.

Panels B and C demonstrate that the faster recovery in the south of England and the Eastern and Western regions is strongly driven by faster growth of online card spending. Notably, within England, the fastest year-on-year growth is in the outer-West area of London, the South West and Eastern England -- areas characterised by highly affluent communities and high levels of second-home ownership. This suggests that, to a degree, spending growth is strongest in the work from home, or potentially work from second home areas of the UK - this is in line with US studies on the implications of COVID-19 for cities \citep[e.g.][]{althoff20,bartik2020jobs,dingel2020many}. In contrast, online spending shows a much slower recovery in the Midlands, North-East and Scotland.

A break-down of the fastest and slowest recovering areas is shown in Table \ref{tab:geography}. The strongest growth in overall spending year-on-year is seen in those regions of the South, East and to the West of London - in and around the ``home counties'' and the affluent second home ownership belt between London and Bristol. In contrast, the slowest growth is seen in Scotland, the North East (Cumbria and Lancashire), the East Midlands (Derbyshire and Nottinghamshire) and West Wales.

\subsection{Variation Across Urban Geographies}

\begin{figure}[H]
	\centering
	\caption{\textbf{Credit Card Spending by Urban-rural Classifications \\ (1 January -  12 December 2020)}}
	\vspace{1cm}
	\begin{tabular}{cc}
		\textbf{A. Affluent England} &
		\textbf{B. Business, Education \& Heritage Sectors} \\
		{\includegraphics[height=1.5in]{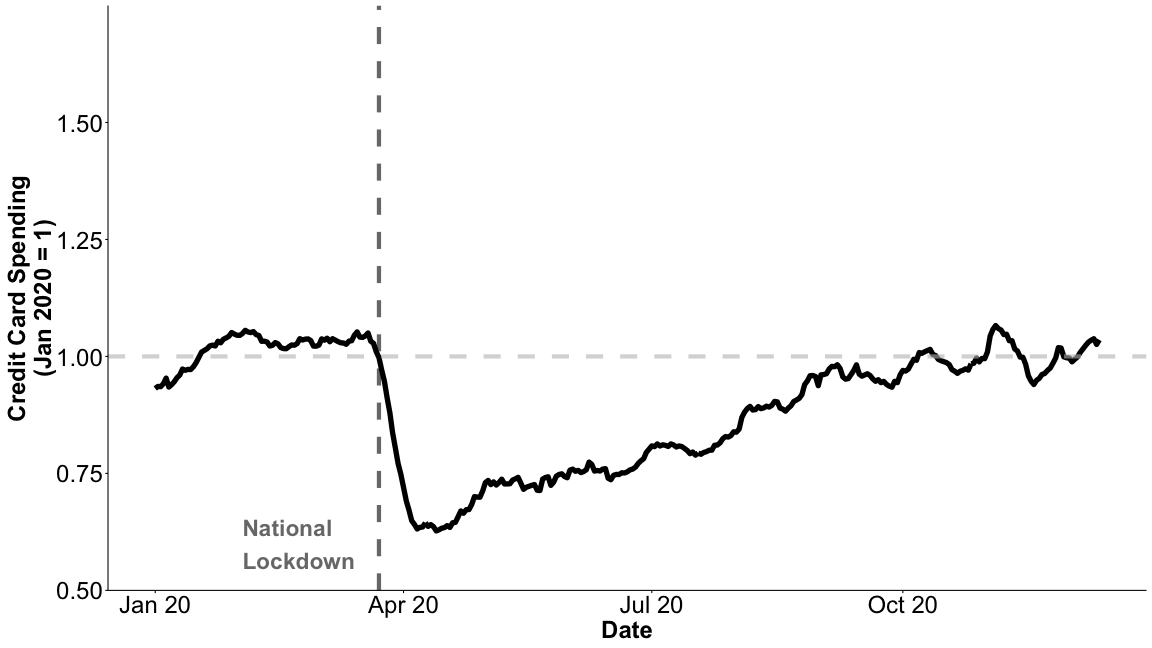}} &  {\includegraphics[height=1.5in]{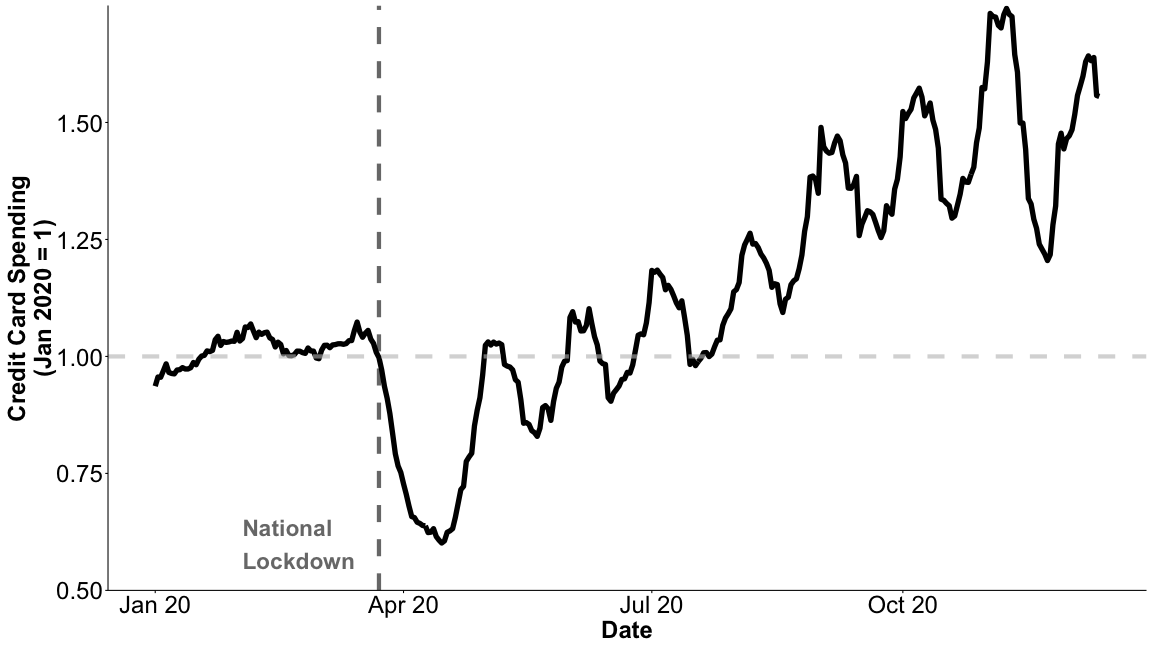}} \\
		\textbf{C. Countryside Living} &
		\textbf{D. Ethnically Diverse Metropolitan Living} \\
		{\includegraphics[height=1.5in]{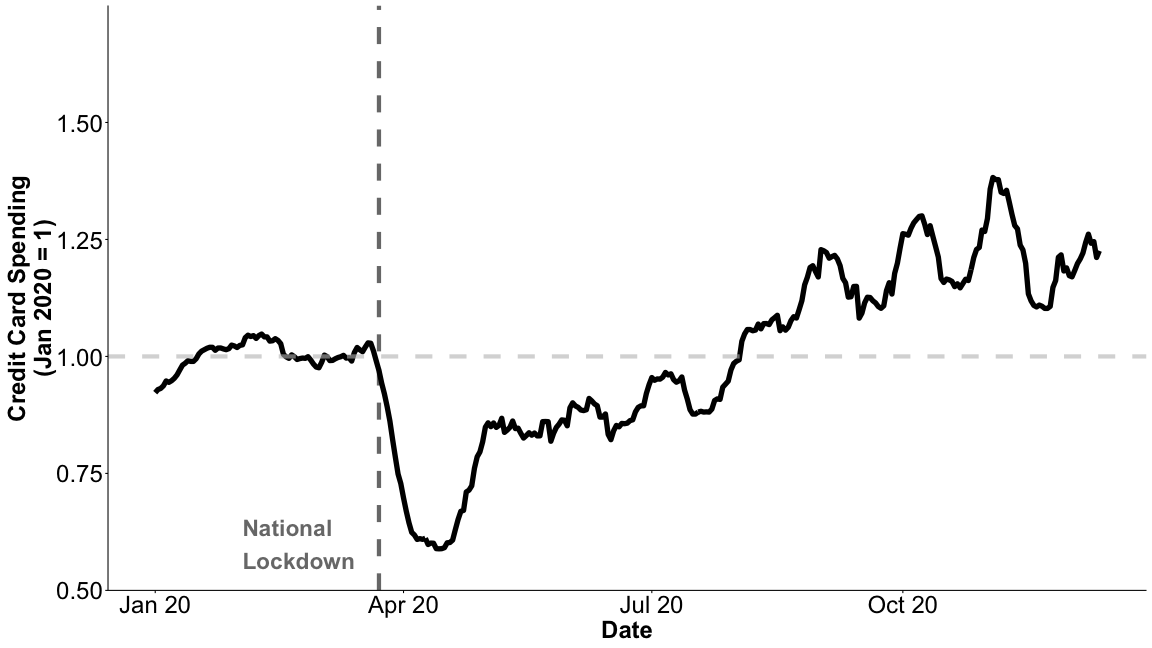}} & {\includegraphics[height=1.5in]{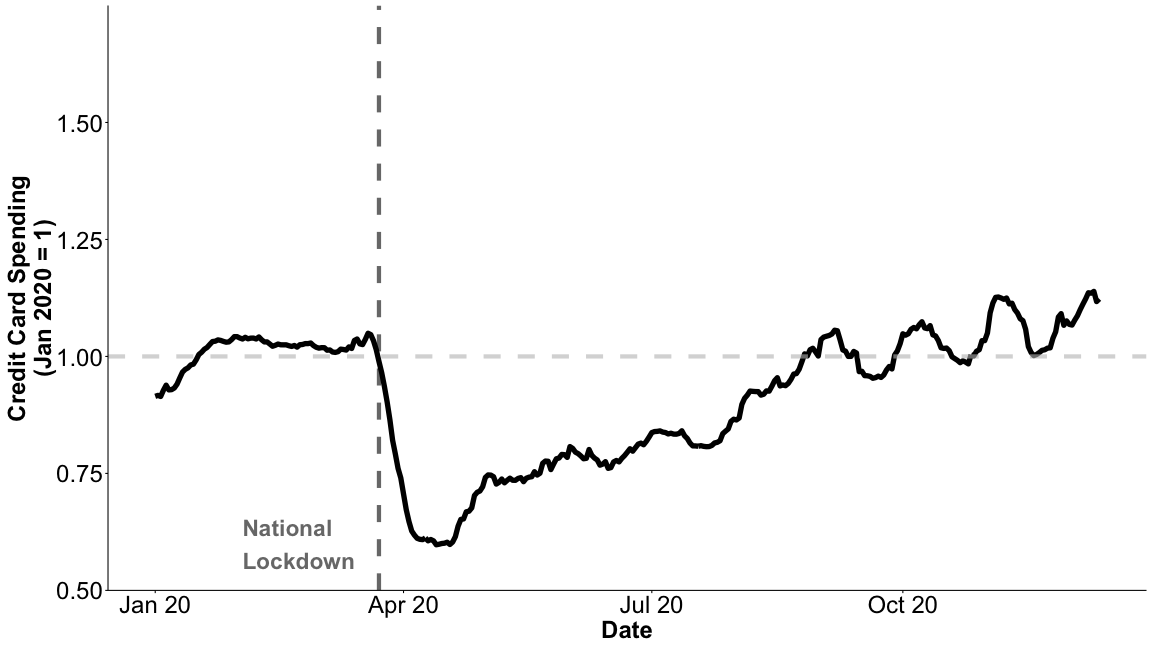}} \\
		\textbf{E. London Cosmopolitan} &
		\textbf{F. Services \& Industrial Legacy} \\
		{\includegraphics[height=1.5in]{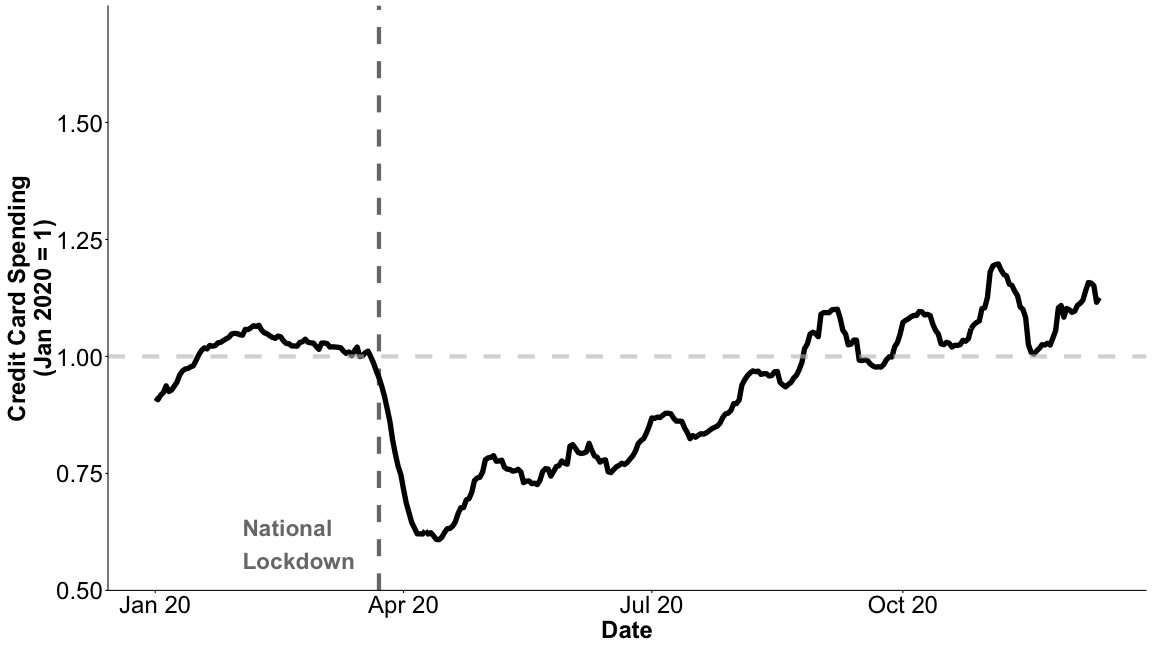}} & {\includegraphics[height=1.5in]{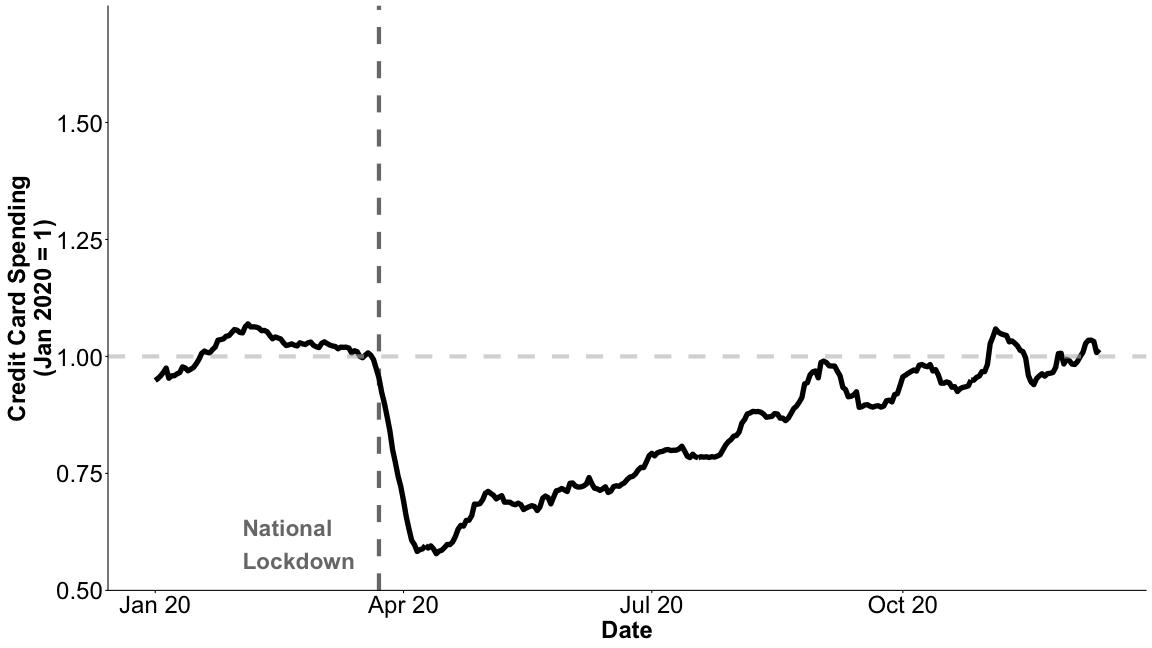}} \\
		\textbf{G. Towns \& Country Living} &
		\textbf{H. Urban Settlements} \\
		{\includegraphics[height=1.5in]{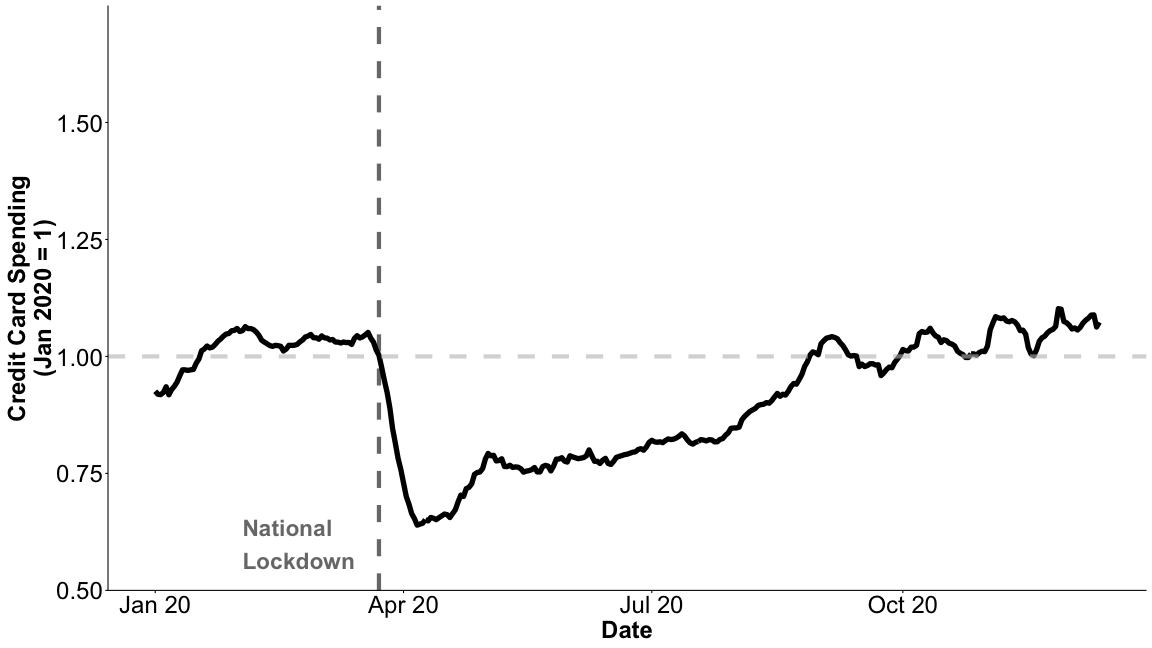}} & {\includegraphics[height=1.5in]{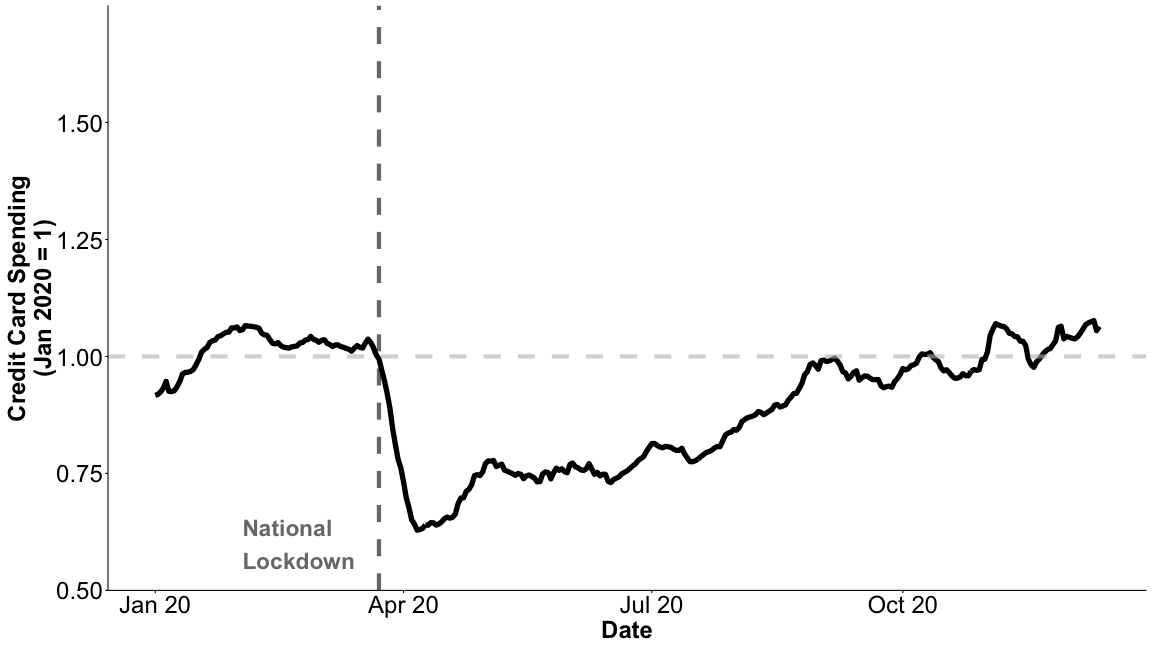}}
	\end{tabular}
	\begin{tablenotes}
		\small
		\item \textit{Notes: Credit card spending is 14 day moving average de-seasoned by taking ratio of the 14 day moving average a year prior. The series is then indexed to its moving average 8 - 28 January 2020. Urban areas presented are the UK official statistics agency the Office for National Statistics (ONS) Super Groups where areas are classified based on the 2011 census. A UK map of these areas can be found in Figure 5 and further details can be found at: \url{https://www.ons.gov.uk/methodology/geography/geographicalproducts/areaclassifications/2011areaclassifications}}
	\end{tablenotes}
	\label{fig:ons}
\end{figure}

To understand the potential of these data further, and the heterogeneous impacts of the COVID-19 crisis, we disaggregate the national series by urban geographies. This updates earlier analyses presented in \cite{gathergood2020ep}.
Figure \ref{fig:ons} disaggregating by eight urban-rural categories created by the UK national statistics agency -- the Office for National Statistics (ONS).\footnote{For maps, methodologies and a detailed description of each category see: \url{https://www.ons.gov.uk/methodology/geography/geographicalproducts/areaclassifications/2011areaclassifications}.}
The figures illustrate that recovery has been fastest and strongest in `business, education and heritage centres' -- such areas are popular domestic tourist destinations and thus this is in line with consumers substituting foreign for domestic holidays.
Recovery was less strong in `countryside living' -- predominately rural areas -- but still noticeably stronger than other, more rural areas.
In more urban areas, `London cosmopolitan' followed by  `ethnically diverse metropolitan living' have had a steady recovery, whereas `affluent England', `services \& industrial legacy', and `urban settlements' are showing weaker recoveries.

\subsection{Variation Across Lockdown ``Tiers''}

Over 2020, the UK has adopted a variety of ``Tier'' systems to impose more or less encroaching limitations on individual movement and social interaction. Following the re-opening after the first national lockdown, a process that began in early May and ended in mid-July, the UK government adopted a policy of ``local lockdowns'', initially on an ad-hoc basis. Particular cities or areas with spikes in positive COVID-19 test rates were subject to these ad hoc restrictions, such as Leicester in the Midlands and Aberdeen in North-East Scotland. The effects of this local lockdown strategy on COVID-19 positive case rates and on card spending are evaluated in \cite{gathergood2020ep}.

In October 2020, the England adopted a formal Tier system, defining three Tiers of restrictions (a similar system was also introduced in Scotland).\footnote{For details, see \url{https://en.wikipedia.org/wiki/First_COVID-19_tier_regulations_in_England}.} These tiers were labelled as ``Medium'', ``High'' and ``Very High''. While most of England was placed in the Tier 1 category, areas around East of London, the Midlands, and areas of the North were categorised into Tiers 2 or 3. Under the rules, in Tier 3 areas certain types of business premises were required to close, while nightclubs and hospitality venues saw their opening hours restricted and their services limited to sit-down meals. Moreover, no mixing between households was permitted indoors or in outdoor private spaces (e.g. gardens), with household mixing allowed only in groups of up to 6 in public outdoor spaces.

By November 2020, the UK government enforced a second national lockdown in England in the form of a four-week period during which pubs, restaurants, leisure centres and non-essential shops would close (in addition to the restrictions in place under Tier 3, as described above).\footnote{Northern Ireland, Scotland and Wales also introduced further restrictions with Table \ref{tab:aggregate_regions}  showing results for these areas.} The UK government re-introduced the job furlough scheme in a form closely resembling that offered in the first national lockdown. This second lockdown ended in early December 2020, with localities being again placed into one of the three Tiers. The allocation of localities to Tiers through 2020 is highly persistent. Localities which have been allocated to Tiers 2 or 3 in October 2020 were all placed in Tier 3 in December 2020. However, a notable difference in the December 2020 classification of areas to Tiers is that very few areas are classified as Tier 1 (these areas together host less than 1.5\% of the UK population).

To provide an indication of the effects of the distribution of the cumulative effects of national and local lockdowns across England, in Table \ref{tab:tiers} we classify NUTS 2 regions by their Tier status as of the start of December 2020 (London, parts of Essex and Hertfordshire was later reclassified into Tier 3 in mid-December and then into a new Tier 4 category on 20 December following an outbreak of a more contagious new, mutant strain of the virus). The tables provide a breakdown of year-on-year growth in overall, offline and online spending for April, July, October, and November.

Table \ref{tab:tiers} shows two main patterns. First, Tier 2 and Tier 3 areas exhibit similar year-on-year growth rates in card spending in April and July 2020 (the period before the Tier system came into operation). However, by October this pattern diverges, with stronger recoveries in overall spending in the Tier 2 areas compared with the Tier 3 areas. This divergence persists through November 2020, with localities facing the new tighter ``Tier 3'' restrictions (mostly the Midlands and Northern Areas) showing 38.4\% lower year-on-year growth in overall spending compared with areas facing the less restrictive ``Tier 2'' (mostly London and the South).
These findings corroborate recent evidence from labour market statistics that the pandemic is levelling down economic activity in the UK, thereby exacerbating regional inequality.
Thus far there is little evidence such lost output of the Tier 3 areas most affected by COVID-19 would naturally recover and so these areas may fall even further behind regions that entered the pandemic with stronger regional economies.

\section{Real-Time Rebalancing Policies}

Addressing regional inequalities that have persisted in the UK and other countries for decades is a hard problem with no quick fixes.
However, the inequalities we document here are more short-term - having developed in 2020.
This offers a potential opportunity for temporary policy interventions targeted at `levelling-up' the regions worst affected by the COVID-19 pandemic to prevent such inequalities becoming persistent.
Such measures could be safely introduced in 2021 once virus outbreaks are deemed to be under control and vaccinations have been more broadly rolled out.

The ability to measure regional economic data in real-time using datasets such as Fable Data offers a groundbreaking potential to inform when, where, and how to target such regional policy interventions.\footnote{To this end, the authors have access to a variety of real-time, high-quality private-sector datasets for research to inform policymaking. If you are a data provider interested in joining this collaboration, please contact the authors for further details on how to potentially partner in this initiative.}
Previously, governments would lack sufficiently real-time data to be able to effectively target interventions in such a rapidly moving crisis.
Even in `normal' times it is extremely difficult to predict the effectiveness of policies but in volatile economic conditions can become even more so.
The ability to evaluate the causal effects of interventions across a breadth of economic outcomes in real-time provides a cost-effective way for governments to trial interventions in selected regions at small scale and quickly consider whether to expand, modify or remove such policies.
This is a more nimble strategy than traditional government approaches of typically applying policies at a national level with limited (if any) ability to assess their impact until it is too late to scale up or down such measures to address the original policy aims.

What could such targeted, regional measures look like? 
We provide a few examples to stimulate public discussions evaluating their relative merits and feasibility of these and other proposals.
Given the clear impacts of offline retailers in particular locations, providing relief to such businesses can help to sustain them and enable them to pass through savings to increase demand for their business.
This could take the form of business rate relief or VAT cuts based on the location of stores.

Given some regions have recovered rather well, providing incentives to encourage residents in those locations to visit harder-hit parts of the UK could be an effective way to generate spending in depressed areas.
One method for doing so would be providing vouchers for discounted travel to and spend in particular destinations ---potentially through a lottery system to enable some individuals to have large incentives to do so.
Such initiatives may also yield broader, longer-term benefits for consumers living in different regions. It may make the labor market more geographically mobile and repair cultural divides.
A less centralised approach would be for national governments to make temporary funding available to local governments.
This could be done in proportion to how adversely such regions have been impacted by the crisis and adjust the duration and amount of funding to local authorities in response to real-time indicators.
Such funding could be allocated to particular uses, however, it may be more efficient to provide flexibility to enable local authorities to use their local knowledge and engage with local residents to use as they consider would be its most productive use.
For example, some areas may decide to use this to provide council tax reductions or rebates to act as a direct form of redistribution.
Other areas may consider forms of spending to be more efficient, such as funding local events or services or initiatives targeted at providing relief to socio-economic groups of people most adversely affected by the pandemic.

\section{Conclusion}

The impact of the COVID-19 pandemic, and the restrictions imposed by the government to limit infections, is geographically uneven. We use newly-available transaction-level credit card data provided by Fable Data to examine geographic variation in the recovery in card spending across the UK. Our analysis shows that the recovery in aggregate spending in the UK masks significant heterogeneity across regions. The recovery is heavily weighted towards the ``home counties'' around outer London and the South, which have shown strong growth in online spending in particular. The strongest recovery in spending is seen in online spending in the ``commuter belt'' areas in outer London and the surrounding localities, and also in areas of high second home ownership, where working from home (including working from second homes) has significantly displaced the location of spending.

We hope by documenting such regional inequalities, our paper helps to provoke informed discussion on this topic and what policy tools could be harnessed to rebalance them in 2021. The availability of real-time data offers the potential for governments to trial a variety of regionally-targeted policies and get real-time feedback on their effectiveness in order to nimbly, expand, modify or remove such measures.
While this paper has focused on the regional impacts of COVID-19, a broad range of other unequal impacts have also developed (e.g. financial and social capital, age, gender, and industry) where partnerships between governments, private sector data providers and academics can help to better measure, understand these impacts and develop well-targeted interventions to attempt to recover lost potential output and rebalance inequalities.
\clearpage

\newpage

\begin{landscape}

\begin{figure}[H]
	\caption{\textbf{April 2020 Card Spend by Geographic Area (Year-on-Year \% Change)}}
	\centering
	\vspace{1cm}
	\begin{subfigure}{0.33\textwidth}
		\centering
		\includegraphics[width=7cm, trim ={17cm 0cm 17cm 0cm},clip ]{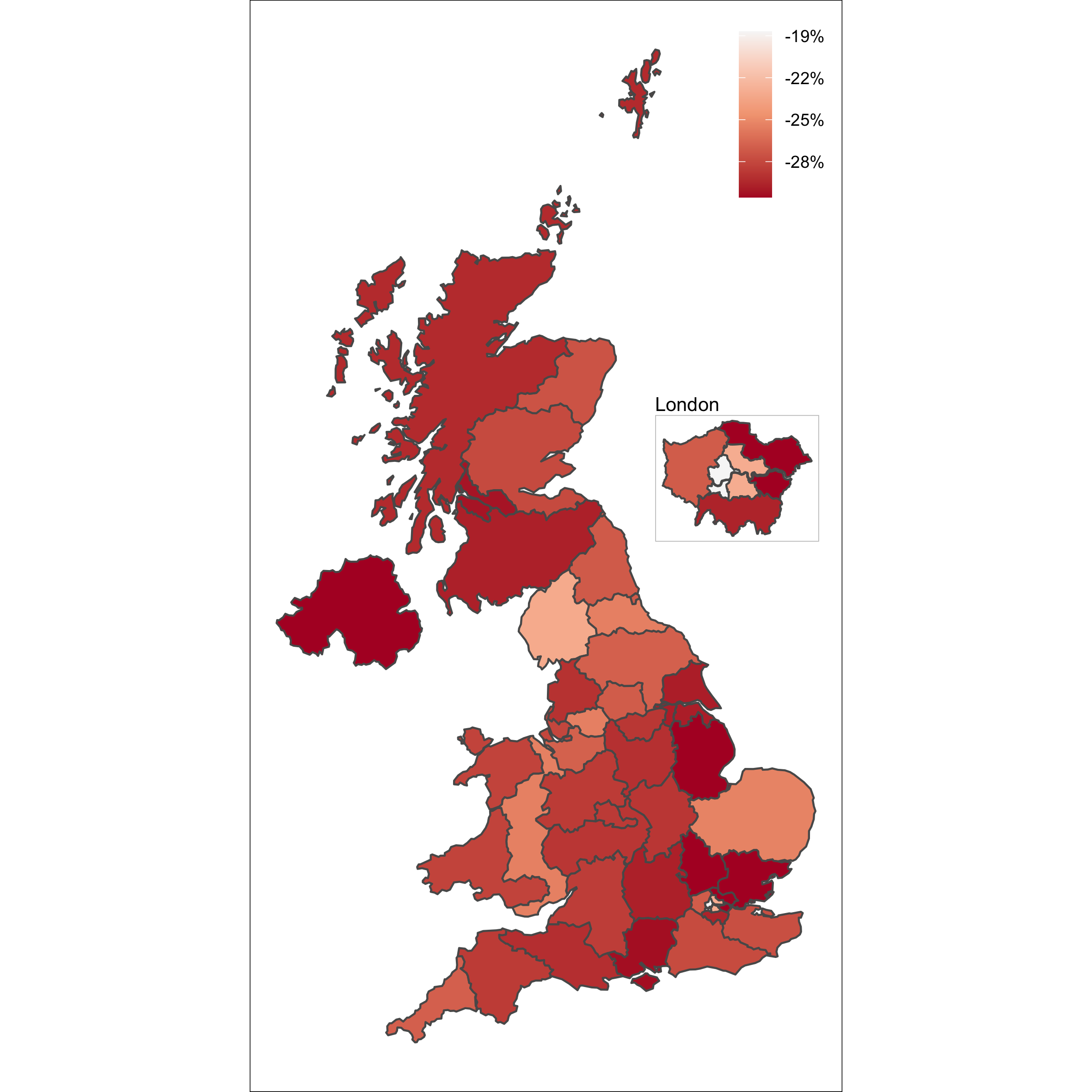}
		\caption{Overall}
	\end{subfigure}
	\hfill
	\begin{subfigure}{0.33\textwidth}
		\centering
		\includegraphics[width=7cm, trim ={17cm 0cm 17cm 0cm},clip]{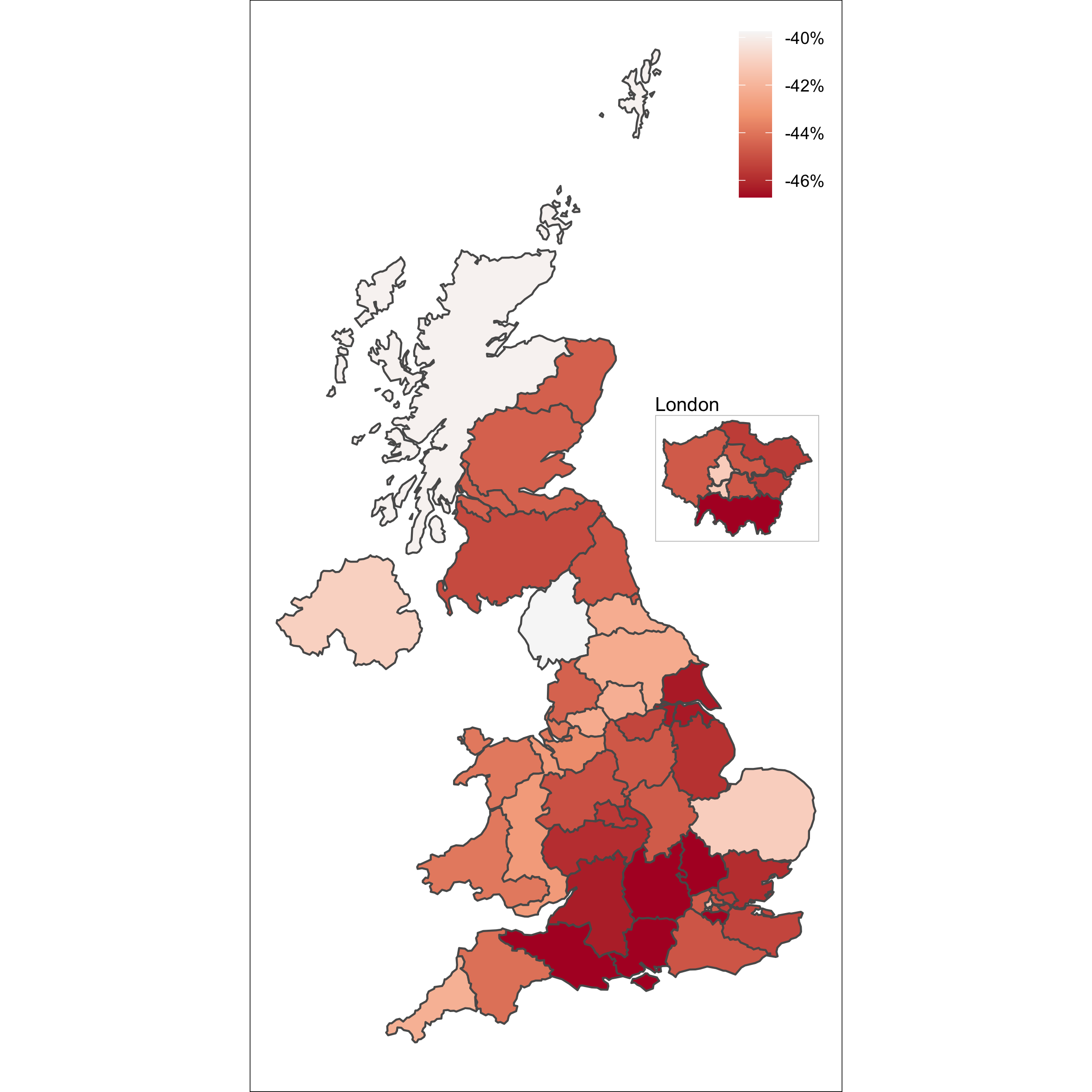}
		\caption{Offline}
	\end{subfigure}
	\hfill
	\begin{subfigure}{0.33\textwidth}
		\centering
		\includegraphics[width=7cm, trim ={17cm 0cm 17cm 0cm},clip]{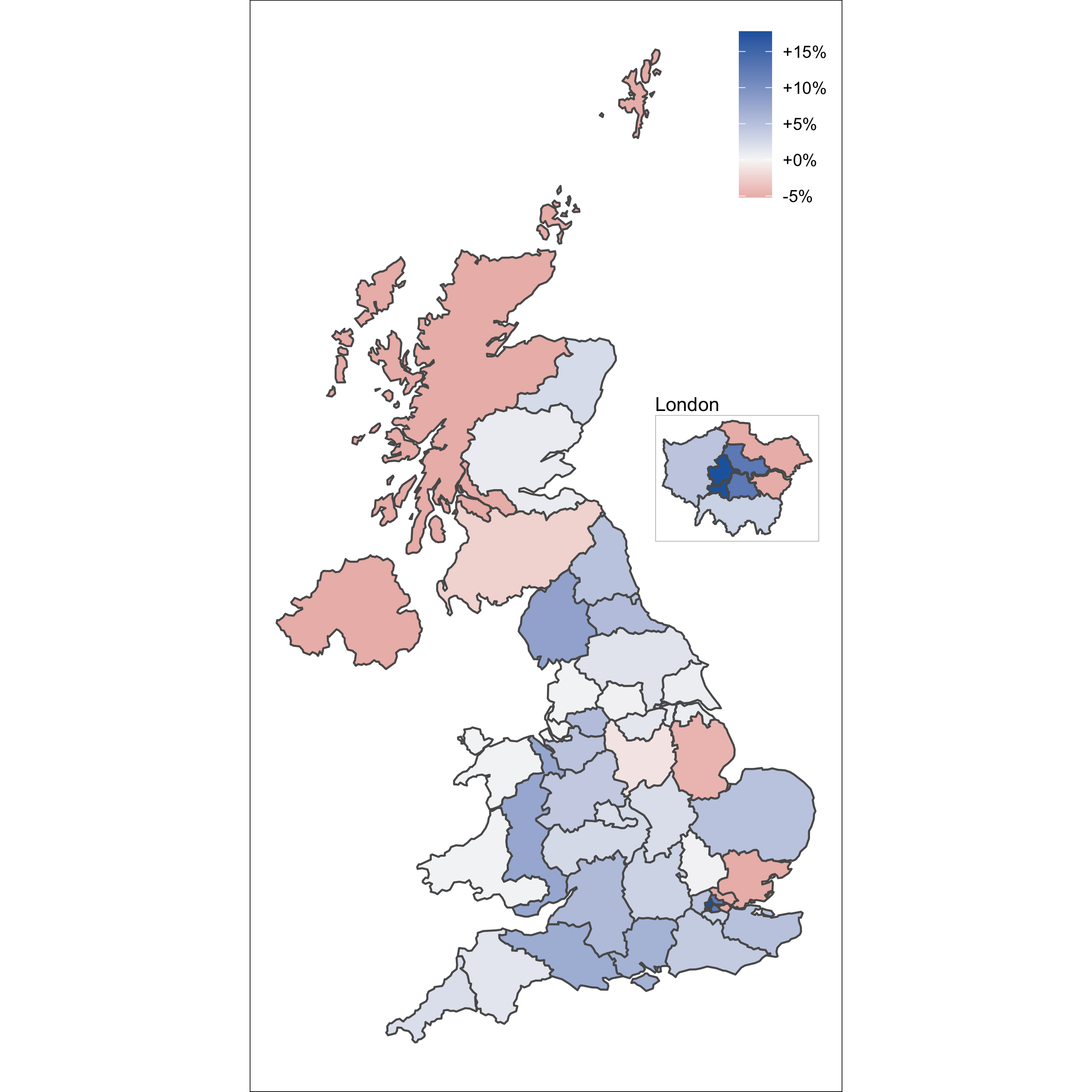}
		\caption{Online}
	\end{subfigure}
		\begin{tablenotes}
		\small
		\item \textit{Notes: Geographic regions are NUTS 2 (Nomenclature of Territorial Units for Statistics) of the United Kingdom.}
	\end{tablenotes}

	\label{fig:map_april}
\end{figure}

\begin{figure}[H]
	\caption{\textbf{October 2020 Card Spend by Geographic Area (Year-on-Year \% Change)}}
	\centering
	\vspace{1cm}
	\begin{subfigure}{0.33\textwidth}
		\centering
		\includegraphics[width=7cm, trim ={17cm 0cm 17cm 0cm},clip ]{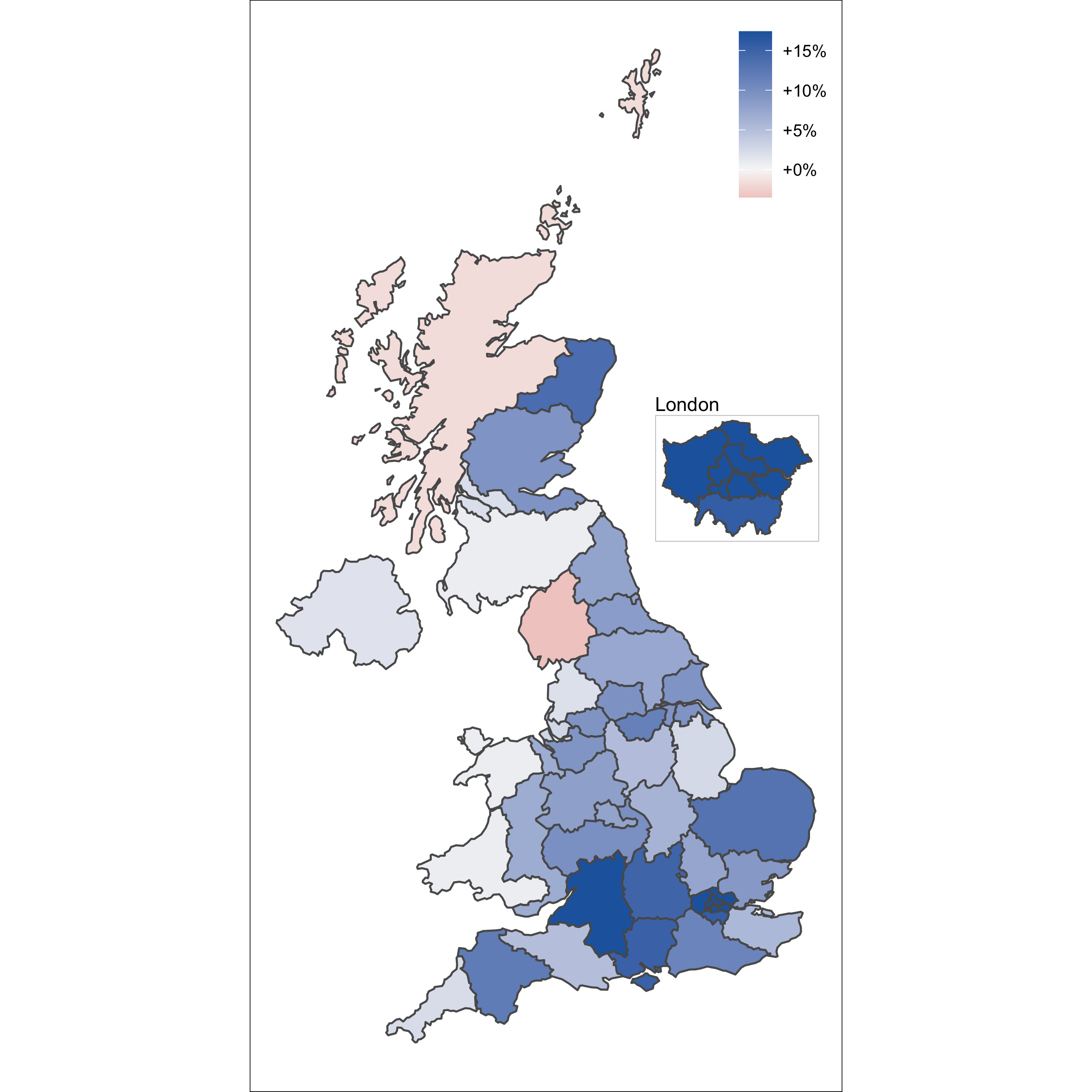}
		\caption{Overall}
	\end{subfigure}
	\hfill
	\begin{subfigure}{0.33\textwidth}
		\centering
		\includegraphics[width=7cm, trim ={17cm 0cm 17cm 0cm},clip]{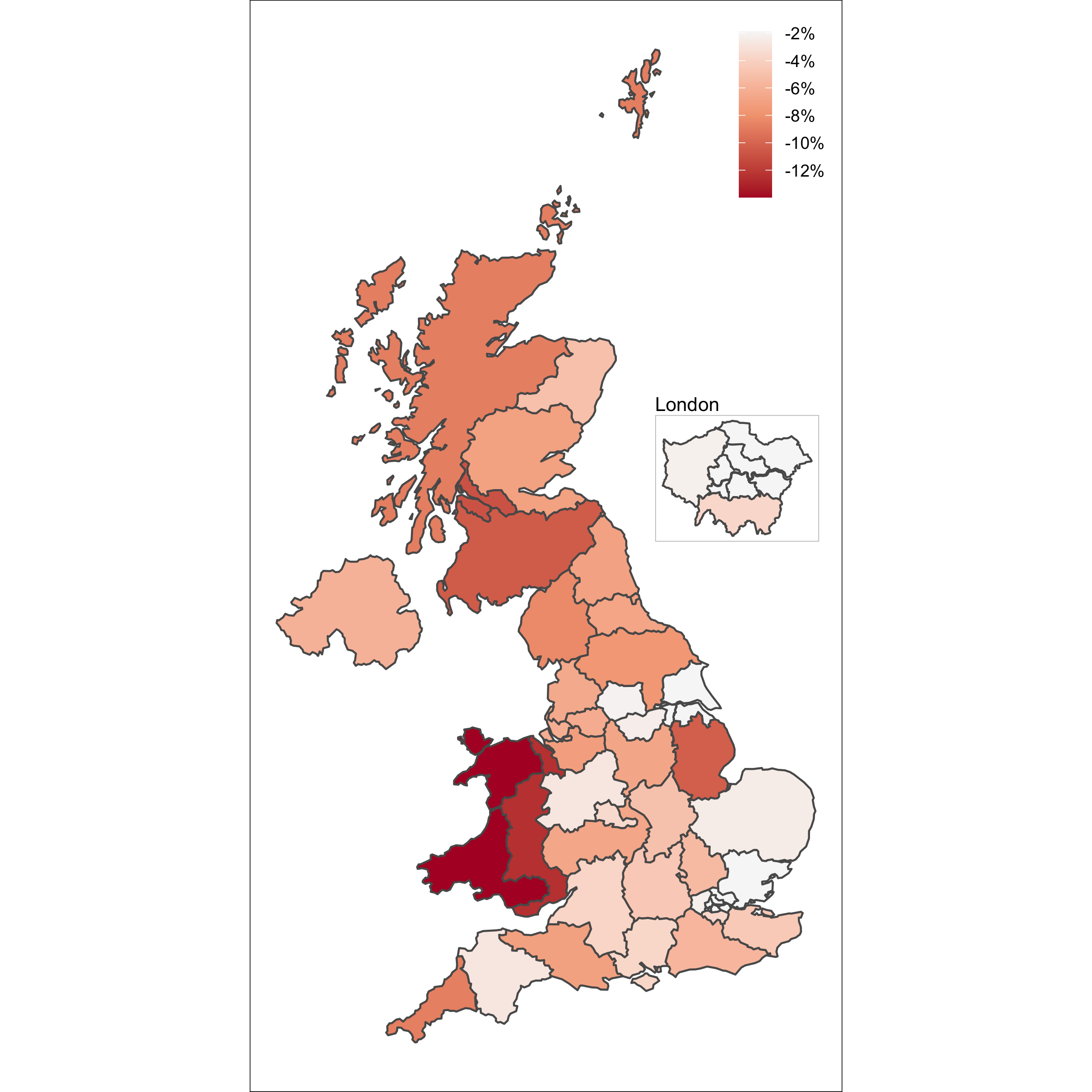}
		\caption{Offline}
	\end{subfigure}
	\hfill
	\begin{subfigure}{0.33\textwidth}
		\centering
		\includegraphics[width=7cm, trim ={17cm 0cm 17cm 0cm},clip]{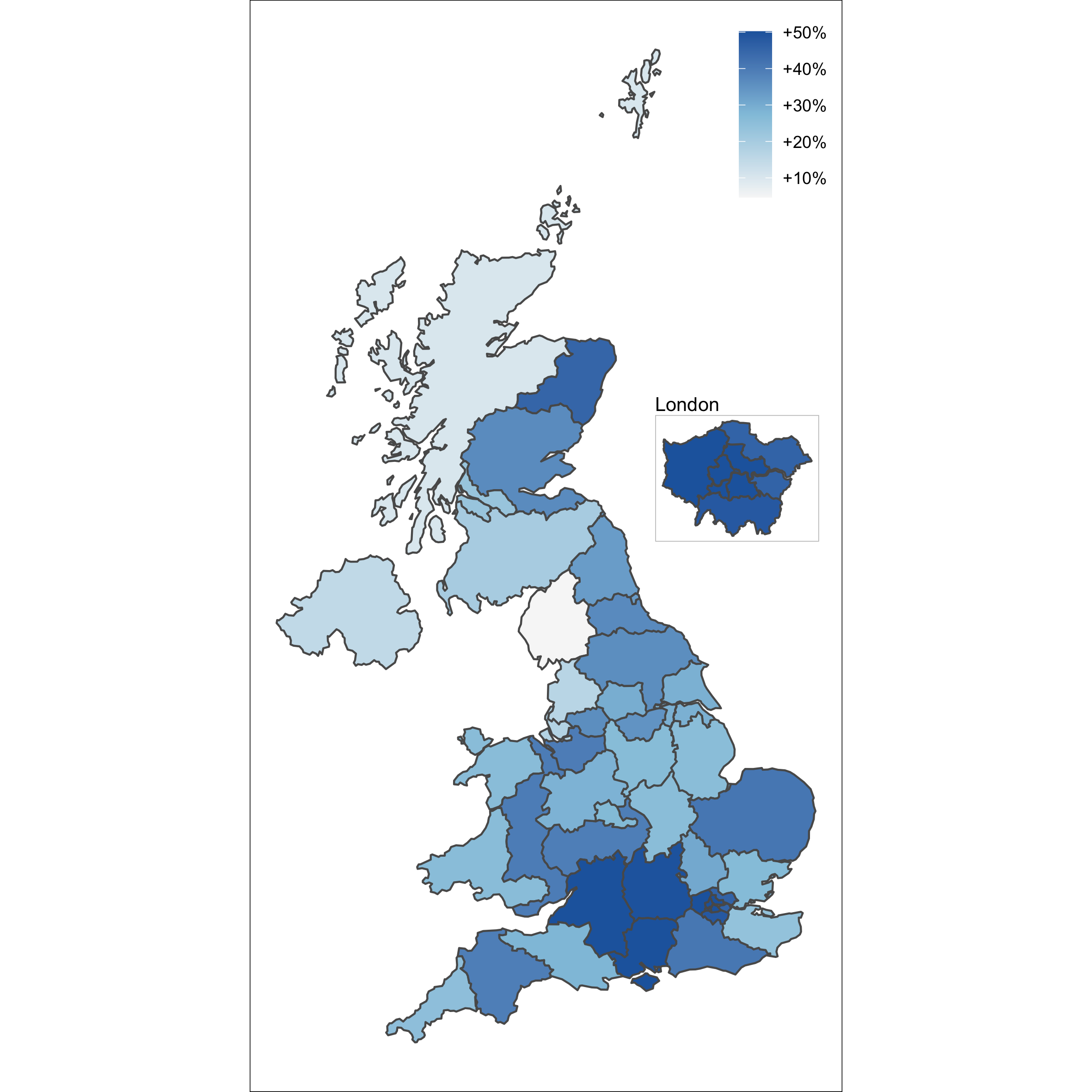}
		\caption{Online}
	\end{subfigure}
	\begin{tablenotes}
		\small
		\item \textit{Notes: Geographic regions are NUTS 2 (Nomenclature of Territorial Units for Statistics) of the United Kingdom.}
	\end{tablenotes}
	\label{fig:map_october}
\end{figure}

\end{landscape}

\newpage

\newpage

\begin{figure}[H]
	\caption{\textbf{UK Urban-rural areas (Super Groups)}}
	\centering
	\vspace{1cm}
	{\includegraphics[width=5in]{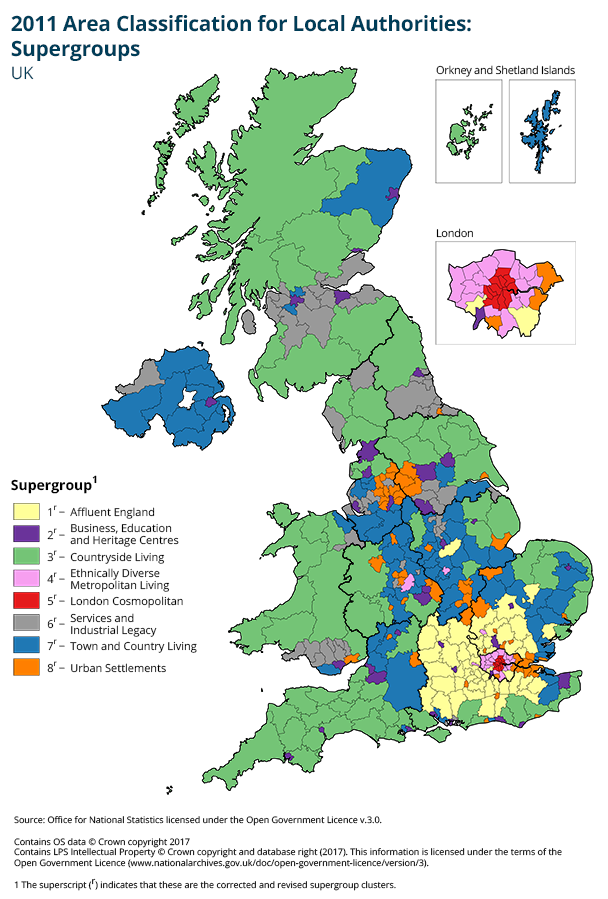}} \\
	\label{fig:mapons}
	\begin{tablenotes}
		\small
		\item \textit{Source: Office for National Statistics (ONS)}
	\end{tablenotes}
\end{figure}
\clearpage
\begin{table}
	\centering
	\caption{\textbf{UK Aggregate Card Spending (Year-on-Year \% Changes)}} 
	\vspace{0.5cm}
	\begin{tabular}{ccccc} \hline
			& \multicolumn{4}{c}{Month}		\\
			\hline
			& April & July & October & November \\
			Overall & -28.1\% & -9.9\% & +11.8\% &  +12.7\% \\
			Offline & -44.7\% & -22.7\% & -4.2\% & -12.0\% \\
			Online & +2.4\% & +12.7\% & + 39.7\% & + 53.1\% \\
			\hline 
	\end{tabular}
	\begin{tablenotes}
		\small
		\item \textit{Notes: Year-on-year change in monthly credit card spending in the UK, including the estimated change in offline and online spending. }
	\end{tablenotes}
	\label{tab:aggregate}
\end{table}

\begin{table}
	\centering
	\caption{\textbf{Aggregate Card Spending Across Regions (Year-on-Year \% Changes)}} 
	\vspace{0.5cm}
	\begin{subtable}[c]{.7\textwidth}
		\centering
		\caption{Northern Ireland}
	\begin{tabular}{ccccc} \hline
		& \multicolumn{4}{c}{Month}		\\
		\hline
		& April & July & October & November \\
Overall & -30.6\% & -13.6\% & 1.6\%  & 10.9\% \\
Offline & -41.0\% & -18.2\% & -6.0\% & -1.1\% \\
Online  & -11.6\% & -5.4\%  & 14.8\% & 30.3\% \\
		\hline 
	\end{tabular}
\vspace{5mm} 
\end{subtable}
	\begin{subtable}[c]{.7\textwidth}
	\centering
	\caption{Scotland}
	\begin{tabular}{ccccc} \hline
		& \multicolumn{4}{c}{Month}		\\
		\hline
		& April & July & October & November \\
Overall & -29.0\% & -15.3\% & 5.0\%  & 9.3\%  \\
Offline & -44.3\% & -26.6\% & -8.9\% & -6.2\% \\
Online  & -3.1\%  & 3.4\%   & 28.0\% & 33.0\% \\
		\hline 
	\end{tabular}
\vspace{5mm} 
\end{subtable}
	\begin{subtable}[c]{.7\textwidth}
	\centering
	\caption{Wales}
	\begin{tabular}{ccccc} \hline
		& \multicolumn{4}{c}{Month}		\\
		\hline
		& April & July & October & November \\
Overall & -27.2\% & -16.0\% & 2.9\%   & 6.0\%  \\
Offline & -43.6\% & -26.6\% & -13.4\% & -9.1\% \\
Online  & 2.9\%   & 3.1\%   & 31.0\%  & 30.4\% \\
		\hline 
	\end{tabular}
\end{subtable}
	\begin{tablenotes}
		\small
		\item \textit{Notes: Year-on-year change in monthly credit card spending across regions, including the estimated change in offline and online spending. }
	\end{tablenotes}
	\label{tab:aggregate_regions}
\end{table}

\begin{table}
	\centering
	\caption{\textbf{Top \& Bottom 5 UK Geographies by Growth in Overall Card Spending \\ (November 2020, Year-on-Year \% Changes)}}
	\vspace{0.5cm}
	\begin{tabular}{cc|cc} \hline
		Top 5 & \% Change & Bottom 5 & \% Change \\
		\hline
		Outer London - South & 16.8\% & Southern Scotland & 3.9\% \\
		Outer London - East & 15.8\% & Cumbria & 3.9\% \\
		Gloucestershire, Wiltshire & 14.6\% & Derbyshire \& Nottinghamshire & 4.3\% \\
		Berkshire, Bucks \& Oxfordshire & 13.5\% & West Wales & 5.2\% \\
		Hampshire \& Isle of Wight & 12.1\% & Lancashire & 5.3\% \\
		\hline
	\end{tabular}
	\begin{tablenotes}
		\small
		\item \textit{Notes: Top and bottom five NUTS 2 areas with the largest annual change in credit card spending during November.}
	\end{tablenotes}
	\label{tab:geography}
\end{table}

\begin{table}
	\centering
	\caption{\textbf{Aggregate Card Spending in England by Dec 2020 Tiers \\ (Year-on-Year \% Changes)}} 
	\vspace{0.5cm}
	\begin{tabular}{ccccc} \hline
		\textbf{Tier 2} & \multicolumn{4}{c}{Month}		\\
		\hline
		& April & July & October & November \\
		Overall & -28.1\% & -9.1\% & +15.6\% &  +15.1\% \\
		Offline & -45.3\% & -22.6\% & -3.2\% & -12.4\% \\
		Online & +3.6\% & +17.2\% & + 47.9\% & + 60.3\% \\
		\hline \\
		\vspace{0.1cm} \\
		\hline
		\textbf{Tier 3} & \multicolumn{4}{c}{Month}		\\
		\hline
		& April & July & October & November \\
		Overall & -27.9\% & -10.1\% & +8.5\% &  +9.3\% \\
		Offline & -44.5\% & -22.4\% & -4.8\% & -14.9\% \\
		Online & +2.3\% & +9.8\% & + 31.9\% & + 49.4\% \\
		\hline		
	\end{tabular}
	\begin{tablenotes}
		\small
		\item \textit{Notes: Year-on-year change in monthly credit card spending by Tier group. }
	\end{tablenotes}
	\label{tab:tiers}
\end{table}

\clearpage

\newpage

\singlespacing 
\bibliography{refs}
\bibliographystyle{apalike}
\doublespacing



\end{document}